# Mastering size and shape of CoPt nanoparticles by flash laser annealing

D. Alloyeau[1,2], C. Ricolleau[1], C. Langlois[1], Y. Le Bouar[2] and A. Loiseau[2]

[1] Laboratoire Matériaux et Phénomènes Quantiques, University Paris 7, 2 Place Jussieu, 75252 Paris, France
[2] Laboratoire d'Etude des Microstructures, ONERA – CNRS, BP 72, 92322 Châtillon, France

A major step towards the understanding of intrinsic properties of nano-objects depends on the ability to obtain assemblies of nanoparticles of a given size with reduced size dispersion and a well defined shape. The control of these parameters is a fundamental challenge. In this newsletter, we present a new method to tailor in an easy way both size and shape characteristics of nanoparticles by using laser irradiation in the nanosecond regime.

The CoPt nanoparticles have been prepared by Pulsed Laser Deposition (PLD) in ultra high vacuum, using a KrF excimer laser at 248 nm with pulse duration of 25 ns. CoPt nanoparticles were formed by an alternative metallic vapor phase deposition on amorphous alumina substrate heated at 550°C. The particles are then covered by a protective thin alumina layer. After the synthesis, the sample was irradiated by using the same laser as the one used for the PLD, but the energy of the laser is well below the ablation threshold of CoPt and $Al_2O_3$ in order to avoid the vaporization of the sample. For that purpose, a focusing lens is placed between the laser and the sample. The energy density on the nanoparticles is controlled by the distance between the back focal plane of the lens and the sample. In our experiment it was fixed to 47 mJ/cm$^2$.

Figure 1a shows the morphology of as-grown CoPt nanoparticles. Due to the tendency of the metals to wet alumina the shape of these particles is elongated in the substrate plane. The morphological changes induced by the laser irradiation as a function of the number of laser pulses is presented in figure 1b and 1c. We observe that this irradiation has two effects. First of all, the morphology of the particles evolves from flat to spherical shape. In the same time, the mean size, the polydispersity (ratio between the standard deviation and the mean size) and the coverage ratio decrease dramatically (table associated to fig. 1).

The mechanisms involved in this experiment can be explained as follow: CoPt nanoparticles absorb the laser light. This absorption results in the increase of the temperature of the nanoparticles and makes possible the desorption of Co and Pt atoms from the particle surface. Due to the increase of the absorption cross section of the UV radiation as the particle size increases,[*] these desorption phenomena are more effective on the biggest particles leading to the reduction of the size and the size dispersion of the particles. In parallel, the laser intensity induces the solid-liquid transition of the alloy leading to a complete reshaping of the particles. This solid-liquid transition is evidenced by the shape of the particles (fig. 1c) similar to small water droplets on clean glass substrate and the formation of twin boundaries characteristic of a solidification process.

CoPt alloy presents two atomic arrangements depending on temperature: a $L1_0$ ordered phase under 825°C and a FCC disordered phase for higher temperature. Same irradiation experiment performed on fully ordered particles induce a phase transformation into the disordered phase, similarly to a quenching of the particles from high temperature phase. CoPt nanoparticles composition is not affected by irradiation.

Flash laser annealing is a very effective way from fundamental and industrial points of view to control size and shape of nanoparticles for the study of their magnetic properties and their applications in high density recording media.

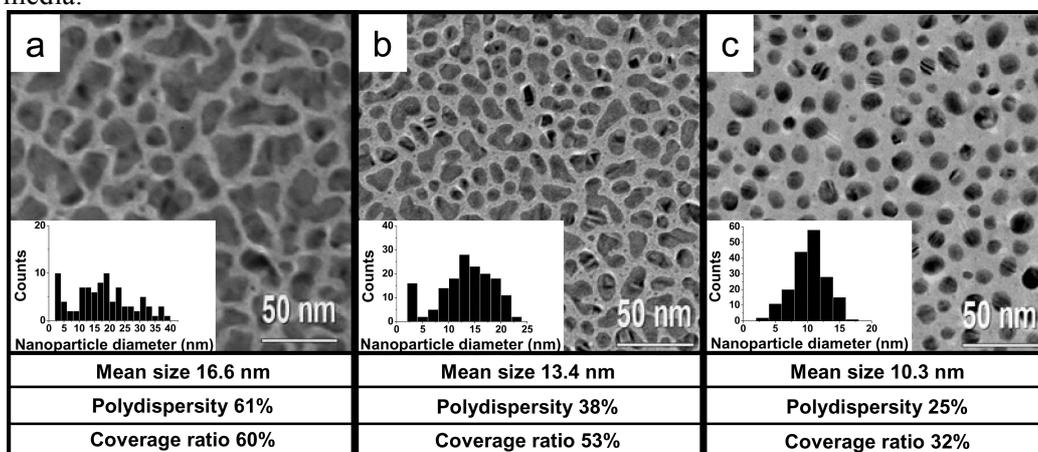

| a | b | c |
|---|---|---|
| Mean size 16.6 nm | Mean size 13.4 nm | Mean size 10.3 nm |
| Polydispersity 61% | Polydispersity 38% | Polydispersity 25% |
| Coverage ratio 60% | Coverage ratio 53% | Coverage ratio 32% |

*Figure 1: TEM images showing the evolution of the particles morphology as a function of the number of laser pulses (fluence of 47 mJ/cm$^2$). (a) As grown nanoparticles. (b) After 1 pulse. (c) After 7 pulses.*

---

[*] J. Bosback, D. Martin, F. Stietz, T. Wenzel and F. Träger, Applied Physics Letters, 1999, 74(18), 2605.